\documentstyle[aps]{revtex}
\begin{document}
\input epsf
\draft
\twocolumn[\hsize\textwidth\columnwidth\hsize\csname
@twocolumnfalse\endcsname
\preprint{PURD-TH-98-06, UH-IfA-98-29, SU-ITP-98-29}
\date{April 1998}
\title{First-order nonthermal phase transition after preheating}
\author{S. Khlebnikov,$^1$  L. Kofman,$^2$  A.
Linde,$^3$ and I. Tkachev$^{1,4}$
 }
\address{
${}^1$Department of Physics, Purdue University, West Lafayette, IN
47907, USA \\
${}^2$Institute for Astronomy, University of Hawaii, 2680 Woodlawn Dr.,
Honolulu, HI 96822, USA \\
${}^3$Department of Physics, Stanford University, Stanford, CA
94305-4060, USA \\
${}^4$Institute for Nuclear Research of the Academy of Sciences of
Russia, Moscow 117312, Russia}
\maketitle
\begin{abstract}
During preheating after inflation, parametric resonance rapidly
generates very large fluctuations of scalar fields.
In models where the inflaton field $\phi$ oscillates in a
double-well potential and interacts with another scalar field $X$,
fluctuations of $X$ can keep the $\phi\to-\phi$ symmetry temporarily
restored.
If the coupling of $\phi$ to $X$ is much stronger than the inflaton
self-coupling, the subsequent symmetry breaking is a
first-order phase transition.
We demonstrate the existence of  this nonthermal  phase transition
with lattice simulations of the full nonlinear
dynamics of the interacting fields. In particular, we observe
nucleation of
an expanding bubble.
\end{abstract}
\vskip2pc]

Cosmological phase transitions are  one of the central topics
of modern cosmology \cite{books}. Recently, this theory was supplemented
by the possibility of {\em nonthermal} cosmological phase transitions
\cite{ptr}, i.e.  phase transitions driven by fluctuations produced so
rapidly that they did not have time to thermalize.
Large nonthermal fluctuations naturally occur in  inflationary
models during preheating  \cite{KLS1}.

Fluctuations of Bose fields generated by the parametric resonance
during preheating have large occupation numbers and  can be considered as
interacting classical waves, which allows to study
the dynamics of fluctuations during and after preheating using lattice
simulations  \cite{KT1}. Numerical calculations, as well as
analytical estimates
\cite{KT1,wide,resc,grav,PR,KLS2,GKLS,khleb} have shown  that
the maximal
values achieved by fluctuations can be large
enough to cause cosmologically interesting phase transitions.
Nevertheless, until now there was no direct demonstration of
the existence of such  phase transitions; several groups which studied
this issue numerically \cite{crit} either concentrated on models where
phase transitions cannot occur, or neglected essential feedback
effects such as
rescattering of created particles.

We have performed a number of lattice simulations of
nonthermal phase transitions, which demonstrated formation of
various types of topological defects \cite{KKLST}.
In this Letter we report results that prove  that nonthermal
phase transitions may take place after preheating even on the scale
as large as the GUT scale $\sim 10^{16}$ GeV. The phase transition
that we
have found is {\em first-order}, which may have particularly important
cosmological implications. First-order phase transitions have a
very clear signature: they proceed through nucleation and subsequent
expansion of a bubble of the new phase inside the old phase. In our
opinion, the presence of this distinctive signature eliminates all
doubts about the possibility of nonthermal phase transitions in the class
of theories under investigation.

As a prototype we will use the model
\begin{equation}
L={1\over 2} (\partial_{\mu}\phi)^2
+{1\over 2} (\partial_{\mu}X)^2 -
{\lambda \over 4}(\phi^{2}-v^{2})^{2}-
{g^{2}\over 2} \phi^{2} X^{2} \; .
\label{L}
\end{equation}
The inflaton scalar field $\phi$ has a double-well potential and
interacts with an $N$-component scalar field $X$;\,
$X^2=\sum_{i=1}^N X_i^2$. For simplicity,
the field $X$ is taken massless and without self-interaction.
The fields couple minimally to gravity in a FRW universe with a scale
factor $a(t)$.

The initial conditions at the beginning of preheating are
determined by the preceding stage of inflation. One can define the
moment when preheating starts
as the moment when the velocity of the field $\phi$ is zero in conformal
time $\eta$, $d\eta=dt/a(t)$, $a(0)=1$.
 This happens when $\phi(0) \approx
0.35 M_{\rm Pl}$
\cite{wide}. We will use rescaled conformal time $\tau=\sqrt{\lambda}
\phi(0)\eta$. Inhomogeneous modes of $\phi$ and all modes of $X$
are taken to be in their conformal vacua at $\tau=0$.
We consider the case $v \alt 10^{-3}
M_{\rm Pl}$,  $g^{2}\gg\lambda$, and take $\lambda=10^{-13}$
\cite{books}. The strength of the resonance depends  nonmonotonically
on  the resonance parameter
$q=g^{2}/4\lambda$, being maximal around $q =n^2/2$
\cite{GKLS}. The condition $g^{2}/\lambda \approx 2n^2 \gg 1$
means that the evolution begins in the regime of a broad parametric
resonance.

The equations of motion for $\phi$ and $X$ are
\begin{eqnarray}
&& {\ddot \phi} + 3 H {\dot \phi} - \nabla^{2}\phi/a^{2}+
\lambda (\phi^{2}-v^{2})\phi
+ g^{2}X^{2}\phi  = 0 \; , \label{eqmphi} \\
&& {\ddot X_i} + 3 H {\dot X_i} - \nabla^{2}X_i/a^{2}+
g^{2}\phi^2 X_i = 0 \; ,
\label{eqmchi}
\end{eqnarray}
where $H=\dot{a}/{a}$.
Substituting $\phi=\phi_{0}+\delta\phi$, where $\phi_{0}$ is the  
inflaton's
homogeneous (``zero'') mode, into (\ref{eqmphi}) and assembling terms
linear in $\phi_{0}$, we obtain the effective mass of $\phi$:
\begin{equation}
m_{\rm eff}^2=-\lambda v^{2} + 3\lambda\langle (\delta\phi)^{2}\rangle
+ g^{2}\langle X^{2} \rangle \; .
\label{eqmzm}
\end{equation}
Angular brackets denote spatial averaging. Because fluctuations rapidly
become classical \cite{KT1}, their variances can be computed as spatial
averages; thus $\langle X^2 \rangle$ is the variance of $X$.
The maximal value of $\langle X^2\rangle$ grows with  $N$, while that of
$\langle (\delta\phi)^{2}\rangle = \langle(\phi-\phi_0)^2\rangle$
does not \cite{resc,khleb}.
We therefore expect that at large $N$ fluctuations
of $X$ will play a more important role in (\ref{eqmzm}) than those
of $\phi$, at least for some time. In general, there is no guarantee
that a
useful effective
potential can be defined for a state far from thermal equilibrium.
We have found  that  in our case Eq. (\ref{eqmzm}) gives a good working
definition for the effective mass.

The idea of nonthermal phase transitions \cite{ptr} is that large
fluctuations of $\langle X^2 \rangle$
(and $\langle (\delta\phi)^2 \rangle$) generated during
preheating can change the shape of the effective potential and lead to
symmetry restoration. Afterwards, the universe expands, the values
of $\langle X^2\rangle$ and $\langle (\delta\phi)^2\rangle$ drop down, and
the phase transition with symmetry breaking occurs.
In the case $g^{2}/\lambda \gg 1$  the phase transition can be of the
first order. This can be expected by analogy with the usual
thermal case \cite{firstorder}.
Let us establish the
{\em necessary} conditions for this transition to occur and
to be of the first order.

(i) At the time of the phase transition, the point $\phi=0$ should
be a local minimum of the effective potential. From (\ref{eqmzm}), we see
that this means that $g^2 \langle X^2 \rangle > \lambda v^2$.

(ii) At the same time, the typical momentum $p_*$ of $X$ particles
should be smaller than $gv$. This is the condition of the existence of a
potential barrier. Particles with momenta $p<gv$ cannot penetrate the
state with $|\phi|\approx v$, so they cannot change the shape of the
effective potential at $|\phi| \approx v$. Therefore, if both conditions
(i) and (ii) are satisfied, the effective potential has a
local minimum at $\phi=0$ and two degenerate  minima at $\phi  
\approx \pm v$.

(iii) Before the minima at $\phi\approx \pm  v$ become deeper than the
minimum
at $\phi=0$, the inflaton's zero mode should decay significantly,
so that it performs small oscillations near $\phi=0$. Then, after
the minimum at $|\phi| \approx v$ becomes deeper than the
minimum at $\phi = 0$, fluctuations of $\phi$ drive the system over the
potential barrier, creating an expanding bubble.

Before performing numerical investigation of   any particular
model one may
want to find out in which cases these conditions can be satisfied.  If the
initial value of $\phi_0$ is much larger than
$v$, the first stages of preheating proceed  in the same way as in the
conformal theories ($v=0$), which is a well established case.
The maximal value of $\langle X^2 \rangle$ is reached at some time
$\tau_{\max}$.
The previous studies \cite{resc,khleb} indicated that
$\langle X^2 \rangle_{\rm max}\sim
\sqrt{N} \phi^2(0)/ q a^2(\tau_{\max})$.
We need, however, an estimate for $\langle X^2 \rangle$ at the time
of the phase transition.
If, at $\tau>\tau_{\max}$, $\langle X^2 \rangle$
merely decreased
with time due to the
redshift, we could  estimate  it by simply replacing
$a(\tau_{\max})$
with the current value of $a$. In reality, fluctuations of $X$ decrease
somewhat slower, because of the continuing decay of the inflaton, but
since the condition (i) holds whenever $v$ is
small enough compared to fluctuations, it is safe to make the replacement
$a(\tau_{\max}) \rightarrow a$.
 As for the condition (ii), we note that the scale of
momenta of $X$ particles is set by the frequency of the inflaton's
oscillations, $p_*\sim \sqrt{\lambda}\phi(0)/a$, although numerical
factors are to be expected in this estimate. Both
conditions (i) and (ii) are met if
$\frac{\phi^2(0)}{a^2 q} \ll v^2 \ll \sqrt{N}  
\frac{\phi^2(0)}{a^2}$. We see
that the window allowed for $v^2$ grows with $q$ and with $N$.
Therefore we will explore cases $g^2/\lambda \gg 1$ and $N\geq 1$.

The condition (iii) cannot be verified using only results
of the previous studies. Nevertheless, we may expect a fairly rapid
decay of the zero mode by analogy with such a decay in the model with
a massive inflaton \cite{resc}. The analogy is relevant when the
amplitude of $\phi_0$ in the present model becomes comparable to $v$,
so that deviations from the conformal invariance set in.
This expectation is well confirmed by our numerical results.

For numerical studies, the full nonlinear equations of motion
(\ref{eqmphi}), (\ref{eqmchi}) were solved directly in the configuration
space.
The computations were done on $64^3$ lattices for $N $= 1, 2, 9 and
on $128^3$ lattices for $N = 1,$  2. Below we present results for
the model with parameters $g^2/\lambda =200$ and
$v = 0.7 \times 10^{-3} M_{\rm Pl} \approx 0.8\times  10^{16}$ GeV, and a
two-component $X$, obtained on a $128^3$ lattice, with the expansion  
of the
universe assumed to be radiation dominated.

\begin{figure}
\leavevmode\epsfysize=5.5 cm \epsfbox{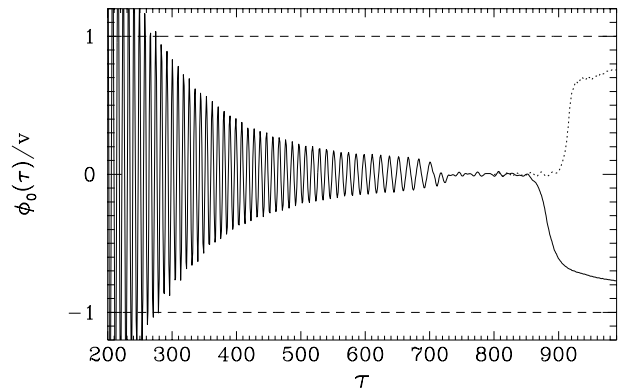}
\caption{Time dependence of the zero-momentum mode of $\phi$ in
units of its vacuum value for two runs with different realizations
of random initial conditions for fluctuations.
All other parameters were the same
in both runs; $\tau$ is conformal time in units of $(\lambda  \phi^2
(0))^{-1/2}$.}
\label{fig:Fig1}
\end{figure}

Time dependence of the zero mode $\phi_0$ is shown in Fig.~\ref{fig:Fig1}.
Initially $\phi_{0}$ oscillates with a large amplitude ${\bar \phi}\gg v$.
If all fluctuations were absent, the zero mode
$\phi_0$, in the expanding universe, would soon start oscillations near
one of its vacuum values, $\pm v$. This would happen when the amplitude
of the oscillations became smaller that $\sqrt{2}v$. In Fig.  
\ref{fig:Fig1} we
see that
the actual dynamics is completely different. The zero mode of the
field  $\phi$ continues to oscillate near $\phi=0$ even when its
amplitude
becomes much smaller than $v$.
In other words, the field oscillates on top of the local maximum of
the bare potential. This can occur only because the effective
potential acquires a minimum at $\phi = 0$ due to interaction
of the field $\phi$ with $\langle X^2\rangle$.

At still later times, $\tau \agt 720$ in Fig. \ref{fig:Fig1}, the
zero mode of $\phi$ decays completely.   This  should  be  
interpreted as
restoration of the symmetry $\phi \to -\phi$    by nonthermal
fluctuations.
Finally, at $\tau \agt 860$ a phase transition occurs, and the
symmetry breaks down.

\begin{figure}
\leavevmode\epsfysize=5.5cm \epsfbox{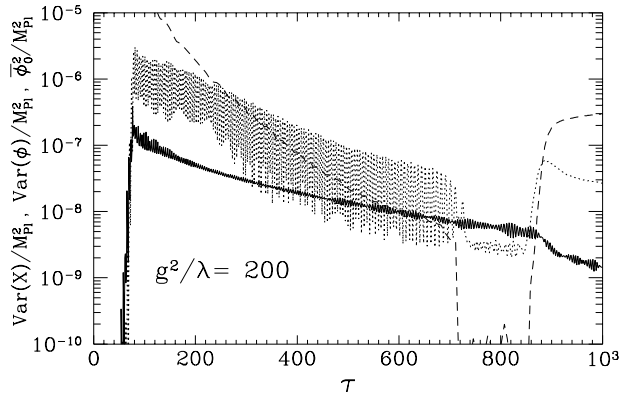}
\caption{Variance of $X$ (solid line), variance of $\phi$
(dotted line), and amplitude of the zero-momentum mode of $\phi$
(dashed line) as functions of time.
}
\label{fig:Fig2}
\end{figure}

Variances of the fields $X$  and $\phi$ as functions of time are shown in
Fig. \ref{fig:Fig2}. One can readily estimate that fluctuations of $X$
(whose variance is multiplied by a larger coupling constant $g^{2}$) give
a dominant contribution to the effective potential, compared to
fluctuations of $\phi$ itself. At the moment of the phase transition
($\tau  \approx 860$ in Fig. \ref{fig:Fig2}), $g^2 \langle X^2 \rangle$
is still larger than $\lambda v^2$. This means that the effective
potential still has a local minimum at $\phi=0$ even after the phase
transition, which confirms that the transition is of the first order.

\begin{figure}
\leavevmode\epsfysize=5.5cm \epsfbox{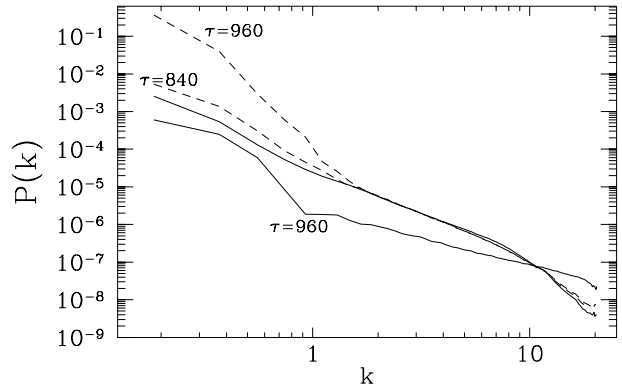}
\caption{Power spectra of the fields $X_1/a$ (solid line) and
$\phi/a$ (dotted line) before and after the phase transition.}
\label{fig:pws}
\end{figure}

To make sure that the phase transition itself is  not a lattice
artifact, we have varied the size $L$ of the integration box and the
number of grid points. Values of $L$ should be chosen in such a
way that, on the one hand, there is enough control over the ultraviolet
part of the spectrum for the results to be cutoff-independent and, on the
other hand, the infrared part of the spectrum is represented well
enough, so
that the order of the phase transition is determined correctly. Good
choices of $L$ are made by monitoring
power spectra of the fields, such as those shown in Fig. \ref{fig:pws}.

At late times (but prior to the phase transition) the power spectra
weakly depend on time and are power law functions at small $k$ with
an exponential tail at large $k$
(conformal momentum $k$ is defined
in units of ${\sqrt{\lambda}} \phi(0)$,
$k \equiv p a/\sqrt{\lambda} \phi(0)$).
 We made sure that the exponential
tail of the power spectra is resolved in our  simulations.
 Under that condition, the power laws do not significantly change
with $L$, the number of grid points and, additionally, with the number
of components of $X$.
The power law for $X$ is fitted by $P_X(k)\propto k^{-2.2}$ for $k > 1$.
At $\tau \approx 700$, $k \approx 10$ corresponds to physical momentum
$p \approx g v$ . Later on, this value
of the physical momentum corresponds to larger $k$,
and at the moment of the phase transition it corresponds to momenta
on the exponential tail of the power spectrum, which signifies that the
condition (ii) is fulfilled, see Fig.~\ref{fig:pws}.
Note the enhancement of power spectra at small $k$ occurring during and
after the phase transition. This is a signature of a bubble of the new
phase and of the ``soft'' fluctuations produced in bubble collisions
\cite{KRT}.

We have found that  the transition grew stronger when we captured more of
the infrared region, either through an increase of the grid size at fixed
maximal momentum or through an increase of $L$ itself. Thus we expect that
the phase transition will be even more strongly first-order if we use
larger lattices for our simulations.

The transition also became more strongly first-order when we increased
$g^2/\lambda$ or the number $N$ of fields $X_i$. Nevertheless, we observed
a first-order transition even at $N=1$.

\begin{figure}
\leavevmode\epsfysize=5.5cm \epsfbox{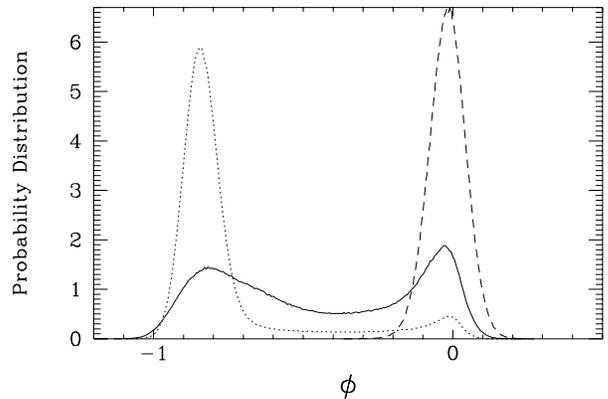}
\caption{Probability distribution of $\phi({\bf x})$ at several
moments of time just before (dashed line), during (solid line),
and at the end of the phase transition (dotted line).}
\label{fig:Fig3}
\end{figure}

\begin{figure}
\leavevmode\epsfysize=6.5cm \epsfbox{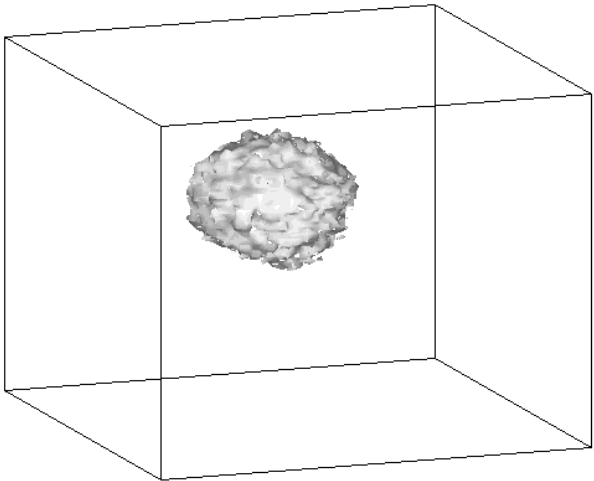}
\caption{Bubble of the new phase. We plot a surface of the constant field
$\phi=-0.7 v$ at the beginning of the phase transition.
Inside the surface $\phi < -0.7 v$.
}
\label{fig:Fig4}
\end{figure}

Direct demonstrations of the first-order nature of the transition are
obtained by plotting a probability distribution of values of $\phi$ over
grid points and the actual field configuration in space. These are
shown in Figs. \ref{fig:Fig3}, \ref{fig:Fig4}.

Prior to the phase transition, we observe a single peak in the probability
distribution function of $\phi$. The peak is  centered at $\phi
\approx 0$,
although its position oscillates slightly, synchronously  with
the oscillations of
the zero mode. The width of the peak is related to the variance
$\langle (\delta \phi)^2\rangle$. At the moment of the phase transition,
a distinctive second  peak appears near the vacuum value of the field
(position of this second peak is shifted somewhat from the vacuum value,
because fluctuations still give a significant contribution to the
effective
potential). After that, the second peak grows, while
the peak at $\phi =0$ decreases and eventually disappears. During
this period,
positions of both peaks do not change. For some realizations of
random initial conditions for fluctuations, the second
peak appeared at $\phi \approx v$; for others, at $\phi \approx -v$.
The time of the phase transition was also changing somewhat with the
realization.
This is an unambiguous signature of spontaneous nucleation of a
bubble of the
new phase in a volume occupied by the old phase.

By plotting the actual field configuration in space, we
have indeed directly observed nucleation of a single bubble and the
bubble's subsequent expansion  until it was occupying the whole
integration volume. The field configuration at the beginning of the phase
transition is shown in Fig.~\ref{fig:Fig4}.  To the best of our
knowledge,
this is the first time when lattice simulations allowed one to see
nucleation of bubbles
during a first-order phase transition.

Let us describe some cosmological implications of our results.  
Models that
exhibit behavior shown in Fig. \ref{fig:Fig1} will lead to domain
structure surviving till present. This conclusion is important
because it allows us to rule out a large class of cosmological models that
lead to domain wall creation.
A different behavior of the zero mode is observed at smaller values of
$g^2/\lambda$ or $N$ (but still $g^2/\lambda \gg 1$).
There, the phase transition occurs when the zero mode
still oscillates near $\phi=0$ with a relatively large amplitude.
This is a new, specifically nonthermal, type of a phase transition.
In such cases, bubbles of +$v$ and --$v$ phases will be nucleated in turn,
but their abundances need not be equal, and for certain values of the
parameters one of the phases may happen not to form  infinite  
domains. Such
models are not ruled out and in fact may have an
interesting observable consequence, an enhanced (by bubble wall  
collisions)
background of relic gravitational waves produced by the mechanism
proposed in Ref. \cite{grav}. In models where the field $\phi$
has many components, the phase transition can lead to creation of strings
or monopoles, instead of domain walls \cite{KKLST}. Finally, if the ratio
$g^2/\lambda$ is sufficiently large, one may expect a short secondary
stage of inflation \cite{ptr}. To investigate this possibility one would
need to study models with $g^2/\lambda \gg 10^2$ or, equivalently,
$g^2 \gg
10^{-11}$, which is quite realistic. However, numerical investigation of
this regime requires lattices of a much greater size than we  currently
use. We hope to return to this issue in a future publication.

This work was supported in part by
DOE grant DE-FG02-91ER40681 (Task B) (S.K. and I.T.),
NSF grants PHY-9219345 (A.L.), PHY-9501458 (S.K. and I.T.),
 AST95-29-225 (L.K. and A.L.), and by the Alfred P. Sloan Foundation  
(S.K.).

\end{document}